\documentclass[aps,nofootinbib,preprintnumbers]{revtex4}
\usepackage{amsmath,amssymb,bm,bbm}

\preprint{IPMU14-0252}

\begin{document}

\title{Hamiltonian structure of scalar-tensor theories beyond Horndeski} 
\author{Chunshan Lin, Shinji Mukohyama, Ryo Namba, Rio Saitou} 
\affiliation{Kavli Institute for the Physics and Mathematics of the
Universe (WPI), Todai Institutes for Advanced Study, The University of
Tokyo, 5-1-5 Kashiwanoha, Kashiwa, Chiba 277-8583, Japan}
\date{\today}

\begin{abstract}
 We study the nature of constraints and the Hamiltonian structure in a
 scalar-tensor theory of gravity recently proposed by Gleyzes, Langlois, 
 Piazza and Vernizzi (GLPV). For the simple case with $A_5=0$, namely
 when the canonical momenta conjugate to the spatial metric are linear
 in the extrinsic curvature, we prove that the number of physical
 degrees of freedom is three at fully nonlinear level, as claimed by
 GLPV. Therefore, while this theory extends Horndeski's scalar-tensor
 gravity theory, it is protected against additional degrees of freedom. 
\end{abstract}

\maketitle

\section{Introduction}

Mysteries in modern cosmology such as inflation, dark energy and dark
matter have been strong motivations for alternative gravity theories
beyond Einstein's general relativity, both in the UV and in the
IR. Because of Lovelock's
theorem~\cite{Lovelock:1971yv,Lovelock:1972vz}, modification of general 
relativity requires inclusion of at least one of the following: (i)
extra degrees of freedom, (ii) extra dimensions, (iii) higher derivative
terms, (iv) extension of (pseudo-)Riemannian geometry, (v)
non-locality. Scalar-tensor theories of gravity are examples of the type
(i).

The most general scalar-tensor theory with three degrees of freedom and
second-order equations of motion was found in 1974 by
Horndeski~\cite{Horndeski:1974wa} and rediscovered recently in the
context of the so-called Galileon
theory~\cite{Nicolis:2008in,Deffayet:2009wt,Deffayet:2011gz}. In this
theory, while each term in the action can in general include more than
two derivatives, the equations of motion are independent of derivatives
higher than second-order. This is achieved by special choice of coupling
constants.

In the context of low-energy effective field theories, one should
include all possible terms that are consistent with symmetries and then
truncate the infinite series of terms according to the standard
derivative expansion and power-counting. In this language, Horndeski's
theory is rather fine-tuned. Such fine-tuning is expected to be detuned
by quantum loops in general.

It is thus of theoretical interest to see what the number of physical
degrees of freedom is in detuned theories. Generic deviation from the
fine-tuning invoked by Horndeski's theory would introduce extra degrees
of freedom, at least formally. If such deviation is small enough then
frequencies or momenta of those extra degrees of freedom are higher than
the cutoff scale of the theory and we can safely integrate them out. The
theory then remains healthy in the domain of its validity as a low
energy effective theory. In this case, although the theory formally (or 
apparently) includes extra degrees of freedom, they are usually
considered unphysical and not included in the physical spectrum of the
theory.

In a recent paper~\cite{Gleyzes:2014dya}, Gleyzes, Langlois, Piazza and
Vernizzi (GLPV) asked a similar but slightly different question: they
asked whether it is possible to extend Horndeski's theory without
introducing extra degrees of freedom even formally, irrespective of
whether they are in the regime of validity of the low energy effective 
theory or not. Considering our complete ignorance of the nature of dark
energy, we consider this as a legitimate attitude. GLPV then proposed a
class of scalar-tensor theories of gravity that extends Horndeski's
theory and claimed that the number of degrees of freedom in this class
of theories remains the same as in Horndeski's theory, i.e. three, at a
fully nonlinear level. However, as we shall see later, their analysis is
not  complete: what is called the momentum constraint in
\cite{Gleyzes:2014dya} lacks a contribution from the scalar degree of
freedom hidden in the lapse function and, as a result, is not 
first-class. The purpose of the present paper is to count the number of
degrees of freedom in the GLPV theory by performing Hamiltonian analysis
properly.

The rest of the paper is organized as follows. In Sec.~\ref{sec:action} 
we briefly describe the action of the GLPV theory in the unitary
gauge, adopting the ADM decomposition. In Sec.~\ref{sec:constraints} we
find the complete set of primary and secondary constraints for
the system and divide them into the set of first-class constraints and
that of second-class constraints. In Sec.~\ref{sec:ndof}, adding
gauge-fixing conditions, we end up with $14$ second-class constraints in
the $20$-dimensional phase space and thus conclude that the number of 
degrees of freedom is three, as claimed by GLPV. Sec.~\ref{sec:summary}
is devoted to a summary and discussions. In Appendix
\ref{app:constraints}, we summarize the Hamiltonian canonical
formulation and the treatment of constraints. Appendix \ref{app:poisson}
outlines the calculations of Poisson brackets.

\section{Unitary gauge action}
\label{sec:action}

As far as the derivative of a scalar field $\partial_{\mu}\phi$ is
timelike, one can choose the time coordinate $t$ so that 
\begin{equation}
 \phi = \phi(t).
\end{equation}
This choice of time coordinate is often called {\it unitary gauge}. By
adopting the ADM decomposition
\begin{equation}
 ds^2 = -N^2dt^2 + h_{ij}(dx^i+N^idt)(dx^j+N^jdt),
\end{equation} 
the action of the GLPV theory in the unitary gauge
is~\cite{Gleyzes:2014dya} 
\begin{equation}
  S = \int d^3x dt N\sqrt{h} \sum_{n=2}^5 L_n,
 \label{action-total}
\end{equation} 
where
\begin{eqnarray}
 L_2 & = & A_2(t,N) \ , \nonumber\\
 L_3 & = & A_3(t,N) K \ , \nonumber\\
 L_4 & = & A_4(t,N) K_2 + B_4(t,N) R \ , \nonumber\\
 L_5 & = & A_5(t,N) K_3 + B_5(t,N) K^{ij}G_{ij} \ ,
\end{eqnarray}
and
\begin{eqnarray}
 K & = & K^i_{\ i} \ , \nonumber\\
 K_2 & = & K^2 - K^i_{\ j}K^j_{\ i} \ , \nonumber\\
 K_3 & = & K^3 - 3KK^{ij}K_{ij} + 2K^i_{\ j}K^j_{\ k}K^k_{\ i} \ .
\end{eqnarray} 
Here, $R$ and $G_{ij}$ are the Ricci scalar and the Einstein tensor of
the $3$-dimensional spatial metric $h_{ij}$, 
\begin{equation}
 K_{ij} = \frac{1}{2N}(\partial_t h_{ij}-D_iN_j-D_jN_i)
\end{equation} 
is the extrinsic curvature, and the spatial indices are lowered and 
raised by $h_{ij}$ and its inverse $h^{ij}$.

\section{Nature of constraints}
\label{sec:constraints}

The action does not include time derivatives of $N^i$ and $N$, and thus
we have the primary constraints
\begin{equation}
 \pi_i = 0, \quad \pi_N = 0, \label{eqn:primary}
\end{equation}
where $\pi_N$ and $\pi_i$ are canonical momenta conjugate to $N$ and
$N^i$, respectively. The canonical momentum conjugate to $h_{ij}$ is 
\begin{equation}
 \pi^{ij} = 
  \frac{\sqrt{h}}{2}
  \left\{ 
   A_3 h^{ij} + 2A_4(h^{ij}K-K^{ij})
   + 3 A_5\left[h^{ij}(K^2-K^k_{\ l}K^l_{\ k})
	  +2(K^i_{\ k}K^{kj}-KK^{ij})\right]
   + B_5G^{ij}\right\}.
 \label{eqn:def-piij}
\end{equation}
The Hamiltonian is then given by 
\begin{equation}
 H = \int d^3x \left(\pi^{ij}\partial_t h_{ij} 
		- N\sqrt{h}\sum_{n=2}^5L_n
		+ \lambda^i \pi_i 
		+ \lambda_N \pi_N \right),
\end{equation}
where $\lambda_N$ and $\lambda^i$ are Lagrange multipliers associated
with the primary constraints (\ref{eqn:primary}). We define the Poisson
bracket as usual by 
\begin{eqnarray}
 \left\{ F, G\right\}_{\rm P} 
  & \equiv &
  \int d^3x 
  \left[ 
   \frac{\delta F}{\delta N(x)}
   \frac{\delta G}{\delta \pi_N(x)}  
   + 
   \frac{\delta F}{\delta N^i(x)}
   \frac{\delta G}{\delta \pi_i(x)}
   +
   \frac{\delta F}{\delta h_{ij}(x)}
   \frac{\delta G}{\delta \pi^{ij}(x)}  
   \right.
   \nonumber\\
 & & 
  \left.
   \qquad\qquad
   -
   \frac{\delta F}{\delta \pi_N(x)}
   \frac{\delta G}{\delta N(x)}  
   - 
   \frac{\delta F}{\delta \pi_i(x)}
   \frac{\delta G}{\delta N^i(x)}
   -
   \frac{\delta F}{\delta \pi^{ij}(x)}
   \frac{\delta G}{\delta h_{ij}(x)}
  \right].
\end{eqnarray}

Since the shift vector $N^i$ enters (\ref{eqn:def-piij}) only implicitly
though the extrinsic curvature $K_{ij}$, we have
\begin{equation}
\left.
 \frac{\delta H}{\delta N^i} 
\right|_{N,h_{ij},\pi_N,\pi_i,\pi^{ij},\lambda_N,\lambda^i}
= 
\left.
 \frac{\delta H}{\delta N^i} 
\right|_{N,h_{ij},\pi_N,\pi_i,K_{ij},\lambda_N,\lambda^i}
= -2\sqrt{h}D_j\left(\frac{\pi^j_{\ i}}{\sqrt{h}}\right),
\end{equation} 
provided that (\ref{eqn:def-piij}) can be solved with respect to $K_{ij}$. Here,
the l.h.s. is the partial functional derivative of $H$, considered as a
$t$-dependent functional of ($N$, $N^i$, $h_{ij}$, $\pi_N$, $\pi_i$,
$\pi^{ij}$, $\lambda_N$, $\lambda^i$), with respect to $N^i$. The second
expression is the partial functional derivative of $H$, considered as a
$t$-dependent functional of ($N$, $N^i$, $h_{ij}$, $\pi_N$, $\pi_i$,
$K_{ij}$, $\lambda_N$, $\lambda^i$), with respect to $N^i$. Hence, the
Hamiltonian is of the following form, 
\begin{equation}
 H = \int d^3x \left({\cal H} + N^i{\cal H}_i
		+ \lambda_N \pi_N + \lambda^i \pi_i \right),
 \label{eqn:Hamiltonian}
\end{equation}
where 
\begin{equation}
 {\cal H}_i \equiv -2\sqrt{h}D_j\left(\frac{\pi^j_{\ i}}{\sqrt{h}}\right), 
  \label{eqn:Hgrav_i}
\end{equation} 
and 
\begin{equation}
 {\cal H} = {\cal H}(t, N, h_{ij}, \pi^{kl})
\end{equation}
depends only on ($t$, $N$, $h_{ij}$, $\pi^{kl}$). Here, $D_j$ is the
$3$-dimensional covariant derivative compatible with the spatial metric
$h_{ij}$.

From now on, we set $A_5=0$ for simplicity. With $A_5=0$, the canonical
momentum conjugate to $h_{ij}$ is 
\begin{equation}
 \pi^{ij} = \frac{\sqrt{h}}{2}
  \left[ A_3 h^{ij} + 2A_4(h^{ij}K-K^{ij}) 
   + B_5G^{ij}\right] \ . 
\end{equation}
Provided that $A_4\ne 0$, this relation is equivalent to
\begin{equation}
 K_{ij} = -\frac{1}{A_4}
   \left[ \frac{1}{\sqrt{h}}
    \left(\pi_{ij}-\frac{1}{2}h_{ij}\pi\right)
    + \frac{A_3}{4}h_{ij}
    - \frac{B_5}{2}\left(R_{ij}-\frac{1}{4}Rh_{ij}\right)
   \right] \ ,
\end{equation}
where $\pi\equiv h_{ij}\pi^{ij}$ and $R_{ij}$ is the Ricci tensor of
$h_{ij}$. Hence, as far as $A_4\ne 0$, i.e. as far as the graviton has a
non-vanishing kinetic term, there is no additional primary constraint
other than (\ref{eqn:primary}). The Hamiltonian is of the form
(\ref{eqn:Hamiltonian})-(\ref{eqn:Hgrav_i}) with 
\begin{eqnarray}
 {\cal H} & = & -N\sqrt{h}
  \left[ \frac{1}{A_4}\left(\frac{\pi^i_{\ j}\pi^j_{\ i}}{h}
		       -\frac{\pi^2}{2h}\right)
   + \frac{A_3\pi}{2\sqrt{h}A_4}
   - \frac{3A_3^2}{8A_4} + A_2 + B_4 R
   \right. \nonumber\\
 & & \left.
      -\frac{B_5}{A_4\sqrt{h}}
      \left(\pi^{ij}R_{ij}-\frac{1}{4}\pi R\right)
      + \frac{A_3B_5}{8A_4}R 
      + \frac{B_5^2}{4A_4}\left(R^{ij}R_{ij}-\frac{3}{8}R^2\right)
	\right] \ .
\end{eqnarray}
Hereafter, we consider ${\cal H}$ as a function of ($t$, $N$, $h_{ij}$,
$\pi^{ij}$).

We have the primary constraints (\ref{eqn:primary}). Since~\footnote{The
first equality in each of the following two equations is weak one since
the Lagrange multipliers in the Hamiltonian may depend on canonical
variables. See (\ref{F1}) for this point. A similar remark applies to
the first equality in each equation in (\ref{eqn:dHtotdt}) and
(\ref{piNCconst}) below.}
\begin{eqnarray}
 \frac{d}{dt}\pi_i(x) &  \approx & \left\{ \pi_i(x), H \right\}_{\rm P} 
  = -{\cal H}_i 
  - \int d^3y
  \left[
   \frac{\delta\lambda^j(y)}{\delta N^i(x)}\pi_j(y)
   + \frac{\delta\lambda_N(y)}{\delta N^i(x)}\pi_N(y)
  \right]
  \approx -{\cal H}_i \ , \nonumber\\
 \frac{d}{dt}\pi_N(x) &  \approx & \left\{ \pi_N(x), H \right\}_{\rm P}
  = -\frac{\partial{\cal H}}{\partial N}
  - \int d^3y
  \left[
   \frac{\delta\lambda^j(y)}{\delta N(x)}\pi_j(y)
   + \frac{\delta\lambda_N(y)}{\delta N(x)}\pi_N(y)
  \right]
  \approx -\frac{\partial{\cal H}}{\partial N} \ , 
\end{eqnarray}
the corresponding secondary constraints are
\begin{equation}
 {\cal H}_i \approx 0 \ , \quad{\cal C} \approx 0 \ , 
\end{equation}
where $\approx$ denotes an equality in the weak sense, i.e. the equality
holds once the constraints are imposed, and we have defined 
\begin{eqnarray}
 {\cal C} & \equiv & - \frac{\partial{\cal H}}{\partial N}
  \nonumber\\
&  = & \sqrt{h}
  \left[ 
   \left(\frac{\pi^i_{\ j}\pi^j_{\ i}}{h}-\frac{\pi^2}{2h}\right)
   \frac{\partial}{\partial N}\left(\frac{N}{A_4}\right)
   + 
   \frac{\pi}{2\sqrt{h}}
   \frac{\partial}{\partial N}\left(\frac{NA_3}{A_4}\right)
   - 
   \frac{3}{8}
   \frac{\partial}{\partial N}\left(\frac{NA_3^2}{A_4}\right)
   + \frac{\partial (NA_2)}{\partial N}
   + R\frac{\partial (NB_4)}{\partial N}
   \right. \nonumber\\
 & & \left.
      - \frac{1}{\sqrt{h}}
      \left(\pi^{ij}R_{ij}-\frac{1}{4}\pi R\right)
      \frac{\partial}{\partial N}\left(\frac{NB_5}{A_4}\right)
      + R\frac{\partial}{\partial N}
      \left(\frac{NA_3B_5}{8A_4}\right)
      + \left(R^{ij}R_{ij}-\frac{3}{8}R^2\right)
      \frac{\partial}{\partial N}
      \left(\frac{NB_5^2}{4A_4}\right) \right] \ .
   \label{eqn:def-calC}
\end{eqnarray}
Since $A_{2,3,4}$ and $B_{4,5}$ depend on $N$, ${\cal C}$ generically
depends on $N$. The constraint ${\cal C}\approx 0$ then determines $N$
if 
\begin{equation}
 \frac{\partial^2{\cal H}}{\partial N^2} \ne 0 \ .
  \label{eqn:ddHdNdNne0}
\end{equation}

It is straightforward to show that~\footnote{The first equality below is
kept weak just in case $f^i$ and/or $g^i$ may depend on canonical
variables, and it becomes strong one if both $f^i$ and $g^i$ are
independent of them. A similar remark applies to the 
first equality in each equation in (\ref{HbarCbar}) and
(\ref{eqn:HtotpiN-HtotC-HtotHtot}) and the second equality in
(\ref{eqn:dHtotdt}) below.}
\begin{equation}
 \left\{ \bar{{\cal H}}[f], \bar{{\cal H}}[g] \right\}_{\rm P}
\approx\bar{{\cal H}}\left[[f,g]\right] \approx 0 \ , 
  \qquad
  \mbox{for }
  {}^{\forall}f^i, \   {}^{\forall}g^i , 
\label{HbarHbar}
\end{equation}
where we have defined
\begin{equation}
 \bar{{\cal H}}[f] \equiv \int d^3x f^i(x){\cal H}_i(x) \ ,\quad
  [f,g]^i \equiv f^j\partial_j g^i - g^j\partial_j f^i \ .
\end{equation} 
However, under the condition (\ref{eqn:ddHdNdNne0}), the Poisson bracket
between ${\cal H}_i(x)$ and ${\cal C}(y)$ fails to vanish weakly
as
\begin{eqnarray}
 \left\{ \bar{{\cal H}}[f], \bar{{\cal C}}[\varphi] \right\}_{\rm P}
  \approx 
  \bar{{\cal C}}[f\partial\varphi] 
  - \int d^3x \frac{\partial^2 {\cal H}}{\partial N^2}
  \varphi f^i\partial_iN
  \approx 
  - \int d^3x \frac{\partial^2 {\cal H}}{\partial N^2}
  \varphi f^i\partial_iN \ ,
  \quad
  \mbox{for }
  {}^{\forall}f^i, \   {}^{\forall}\varphi \ ,
\label{HbarCbar}
\end{eqnarray} 
where we have defined
\begin{equation}
 \bar{{\cal C}}[\varphi] \equiv \int d^3x \varphi(x){\cal C}(x) \ ,
  \quad
  f\partial\varphi \equiv f^i\partial_i\varphi \ . 
\end{equation} 
Therefore, contrary to what was claimed by GLPV~\cite{Gleyzes:2014dya},
the constraint ${\cal H}_i\approx 0$ is not first-class.

Nonetheless, defining the following linear combination of constraints
\begin{equation}
 {\cal H}_i^{\rm tot} = {\cal H}_i + \pi_N \partial_i N \ ,
  \label{eqn:def-Htot_i}
\end{equation}
it is possible to show that
\begin{eqnarray}
 \left\{ \bar{{\cal H}}^{\rm tot}[f], \bar{\pi}_N[\varphi] \right\}_{\rm P}
  & \approx & \bar{\pi}_N[f\partial\varphi] \approx 0 \ , \nonumber\\
 \left\{ \bar{{\cal H}}^{\rm tot}[f], \bar{{\cal C}}[\varphi] \right\}_{\rm P}
  &\approx& \bar{{\cal C}}[f\partial\varphi] \approx 0 \ , \nonumber\\
 \left\{ \bar{{\cal H}}^{\rm tot}[f], 
  \bar{{\cal H}}^{\rm tot}[g] \right\}_{\rm P}
  &\approx & \bar{{\cal H}}^{\rm tot}\left[[f,g]\right] \approx 0 \ , 
  \qquad
  \mbox{for }
  {}^{\forall}f^i, \   {}^{\forall}g^i, \   {}^{\forall}\varphi \ ,
  \label{eqn:HtotpiN-HtotC-HtotHtot}
\end{eqnarray} 
where we have defined
\begin{equation}
 \bar{{\cal H}}^{\rm tot}[f] \equiv 
  \int d^3x f^i(x){\cal H}^{\rm tot}_i(x) \ ,\quad
 \bar{\pi}_N[\varphi] \equiv \int d^3x \varphi(x)\pi_N(x) \ .
\end{equation} 
By definition we also have 
\begin{equation}
  \left\{ \pi_i(x), \pi_j(y) \right\}_{\rm P} = 0 \ , \quad
  \left\{ \pi_i(x), \pi_N(y) \right\}_{\rm P} = 0 \ , \quad
   \left\{ \pi_i(x), {\cal C}(y) \right\}_{\rm P} = 0 \ , \quad 
   \left\{ \pi_i(x), {\cal H}^{\rm tot}_j(y) \right\}_{\rm P} = 0 \ .
\end{equation}
Furthermore, 
\begin{equation}
 \frac{d}{dt}\bar{{\cal H}}^{\rm tot}[f]
\approx
  \left\{\bar{{\cal H}}^{\rm tot}[f], H\right\}_{\rm P} 
\approx
  \bar{{\cal H}}\left[[f,N]\right]
  + \bar{\pi}_N[f\partial\lambda_N]
  \approx 0 \ . \label{eqn:dHtotdt}
\end{equation}
Therefore, it is concluded that constraints $\pi_i\approx 0$ and 
${\cal H}^{\rm tot}_i\approx 0$ ($i=1,2,3$) are first-class and that
there is no additional secondary constraint associated with them.

It is easy to show that
\begin{eqnarray}
 \left\{ \pi_N(x), \pi_N(y) \right\}_{\rm P}  & = & 0 \ , \nonumber\\
 \left\{ {\cal C}(x), \pi_N(y) \right\}_{\rm P}  & = & 
  -\frac{\partial^2{\cal H}}{\partial N^2} \delta^3(x-y) \ .
\end{eqnarray}
Hence, provided that the condition (\ref{eqn:ddHdNdNne0}) is satisfied,
the determinant
\begin{equation}
\det
 \left(
   \begin{array}{cc}
    \left\{\pi_N(x), \pi_N(y)\right\}_{\rm P} & 
     \left\{\pi_N(x), {\cal C}(y)\right\}_{\rm P} \\
    \left\{{\cal C}(x), \pi_N(y)\right\}_{\rm P} & 
     \left\{{\cal C}(x), {\cal C}(y)\right\}_{\rm P}
   \end{array}
       \right)
\end{equation} 
does not vanish weakly, meaning that the set of constraints
$\pi_N\approx 0$ and ${\cal C}\approx 0$ is second-class.

The total Hamiltonian is 
\begin{eqnarray}
 H_{\rm tot} & = & \int d^3x 
  \left[{\cal H} + N^i{\cal H}_i + n^i{\cal H}^{\rm tot}_i 
   + \lambda^i \pi_i + \lambda_N\pi_N + \lambda_{{\cal C}}{\cal C} 
  \right]  \nonumber\\
 & = & \int d^3x 
  \left[{\cal H} + (N^i+n^i){\cal H}_i + \lambda^i \pi_i 
   + (\lambda_N+n^i\partial_iN) \pi_N + \lambda_{{\cal C}}{\cal C} \right] \ , 
\end{eqnarray} 
where $n^i$ and $\lambda_{{\cal C}}$ are Lagrange multipliers. Since the
set of constraints $\pi_N\approx 0$ and ${\cal C}\approx 0$ is
second-class, the consistency conditions, 
\begin{equation}
 \frac{d}{dt}\pi_N(x) \approx
  \left\{\pi_N(x), H_{\rm tot}\right\}_{\rm P} \approx 0 \ , \quad
  \frac{d}{dt}{\cal C}(x) \approx
  \frac{\partial}{\partial t}{\cal C}(x) 
  + \left\{{\cal C}(x), H_{\rm tot}\right\}_{\rm P} \approx 0 \ ,
\label{piNCconst}
\end{equation}
determine the two Lagrange multipliers $\lambda_N$ and $\lambda_{{\cal
C}}$, instead of generating additional secondary constraints.

\section{Number of degrees of freedom}
\label{sec:ndof}

One can fix the gauge freedom associated with the first-class
constraints $\pi_i\approx 0$ and ${\cal H}^{\rm tot}_i\approx 0$ by
imposing additional conditions 
\begin{equation}
 {\cal G}^i(x)\approx 0 \ , \quad {\cal F}^i(x) \approx 0 \ , \quad (i=1,2,3) \ ,
\end{equation}
provided that the determinant
\begin{equation}
 \det 
  \left(
   \begin{array}{cc}
    \frac{\delta{\cal G}^j(y)}{\delta N^i(x)} &
     \frac{\delta{\cal F}^j(y)}{\delta N^i(x)} \\
    \left\{{\cal H}^{\rm tot}_i(x), {\cal G}^j(y)\right\}_{\rm P} &
     \left\{{\cal H}^{\rm tot}_i(x), {\cal F}^j(y)\right\}_{\rm P}
   \end{array}
   \right)
\end{equation} 
does not vanish weakly. Including the gauge fixing conditions, we thus
have the following set of $14$ second-class constraints:
\begin{equation}
 {\cal H}^{\rm tot}_i \approx 0, \quad  \pi_i \approx 0 \ , \quad 
 {\cal G}^i(x)\approx 0, \quad {\cal F}^i(x) \approx 0 \ , \quad
 \pi_N \approx 0, \quad {\cal C}\approx 0 \ , \quad (i=1,2,3) \ .
 \label{eqn:14constraints}
\end{equation}
The total Hamiltonian after gauge fixing is thus
\begin{eqnarray}
 H'_{\rm tot} & = & \int d^3x 
  \left[{\cal H} + N^i{\cal H}_i + n^i{\cal H}^{\rm tot}_i 
   + \lambda^i \pi_i + \lambda^{{\cal G}}_i{\cal G}^i
   + \lambda^{{\cal F}}_i{\cal F}^i
   + \lambda_N\pi_N + \lambda_{{\cal C}}{\cal C} 
	 \right] \nonumber\\
 & = &  \int d^3x 
  \left[{\cal H} + (N^i+n^i){\cal H}_i + \lambda^i \pi_i 
   + \lambda^{{\cal G}}_i{\cal G}^i
   + \lambda^{{\cal F}}_i{\cal F}^i
   + (\lambda_N+n^i\partial_iN) \pi_N + \lambda_{{\cal C}}{\cal C}
	 \right], 
\end{eqnarray} 
where $\lambda^{{\cal G}}_i$ and $\lambda^{{\cal F}}_i$ are Lagrange 
multipliers. As usual with the second-class constraints, the set of all
Lagrange multipliers ($n^i$, $\lambda^i$, $\lambda^{{\cal G}}_i$,
$\lambda^{{\cal F}}_i$, $\lambda_N$, $\lambda_{{\cal C}}$) are fully
determined by imposing 
\begin{eqnarray}
&& \left\{{\cal H}_i(x), H'_{\rm tot}\right\}_{\rm P} \approx 0 \ , \quad
 \left\{\pi_i(x), H'_{\rm tot}\right\}_{\rm P} \approx 0 \ , \quad
 \frac{\partial}{\partial t}{\cal G}^i(x) + 
 \left\{{\cal G}^i(x), H'_{\rm tot}\right\}_{\rm P} \approx 0 \ , \quad
 \frac{\partial}{\partial t}{\cal F}^i(x) + 
 \left\{{\cal F}^i(x), H'_{\rm tot}\right\}_{\rm P} \approx 0 \ , \nonumber\\
&&
 \left\{\pi_N(x), H'_{\rm tot}\right\}_{\rm P} \approx 0 \ , \quad
 \frac{\partial}{\partial t}{\cal C}(x) + 
 \left\{{\cal C}(x), H'_{\rm tot}\right\}_{\rm P} \approx 0 \ .
\end{eqnarray}
Hence, starting with the $20$-dimensional phase space ($N$, $N^i$, 
$h_{ij}$, $\pi_N$, $\pi_i$, $\pi^{ij}$), we end up with $6$-dimensional 
physical phase space after imposing the $14$ second-class 
constraints (\ref{eqn:14constraints}). Therefore, the number of degrees
of freedom is three, as claimed by GLPV.

As a simple example of the gauge fixing functions, let us consider 
\begin{equation}
 {\cal G}^i = N^i \ , \quad 
  {\cal F}^i = {\cal F}^i(N, h_{ij}, \pi_N, \pi^{kl};t) \ ,
\end{equation}
such that the determinant
\begin{equation}
\det
 \left(
 \left\{{\cal H}^{\rm tot}_i(x), {\cal F}^j(y)\right\}_{\rm P}
 \right)
\end{equation}
does not vanish weakly. In this case the consistency conditions
\begin{equation}
 \frac{d{\cal G}^i}{dt} \approx 0 \ , \quad 
  \frac{d\pi_i}{dt} \approx  0 \ ,
\end{equation}
determine the six Lagrange multipliers $\lambda^i$ and 
$\lambda^{{\cal G}}_i$ as
\begin{equation}
 \lambda^i = 0 \ , \quad \lambda^{{\cal G}}_i = -{\cal H}_i \ .
\end{equation}
By substituting them, we obtain
\begin{equation}
 H'_{\rm tot} = \int d^3x 
  \left[{\cal H} + n^i{\cal H}^{\rm tot}_i 
   + \lambda^{{\cal F}}_i{\cal F}^i
   + \lambda_N \pi_N + \lambda_{{\cal C}}{\cal C}
	 \right].  \label{eqn:H'tot_gaugefixed}
\end{equation}
Together with the gauge fixing condition ${\cal G}^i=N^i\approx 0$, we
see that the canonical pair ($N^i$, $\pi_i$) is fully eliminated from
the phase space. The dimension of the reduced phase space ($N$,
$h_{ij}$, $\pi_N$, $\pi^{ij}$) is $14$. As usual with the second-class
constraints, the set of all remaining Lagrange multipliers ($n^i$,
$\lambda^{{\cal F}}_i$, $\lambda_N$, $\lambda_{{\cal C}}$) are fully
determined by imposing 
\begin{eqnarray}
&& \left\{{\cal H}^{\rm tot}_i(x), H'_{\rm tot}\right\}_{\rm P} \approx 0, \quad
 \frac{\partial}{\partial t}{\cal F}^i(x) + 
 \left\{{\cal F}^i(x), H'_{\rm tot}\right\}_{\rm P} \approx 0, \nonumber\\
&&
 \left\{\pi_N(x), H'_{\rm tot}\right\}_{\rm P} \approx 0, \quad
 \frac{\partial}{\partial t}{\cal C}(x) + 
 \left\{{\cal C}(x), H'_{\rm tot}\right\}_{\rm P} \approx 0.
\end{eqnarray}
There remains the following set of $8$ second-class constraints acted on
the $14$-dimensional reduced phase space: 
\begin{equation}
 {\cal H}^{\rm tot}_i \approx 0, \quad  
 {\cal F}^i(x) \approx 0, \quad
 \pi_N \approx 0, \quad {\cal C}\approx 0, \quad (i=1,2,3).
\end{equation}
Hence, we end up with $6$-dimensional physical phase space, and the
number of degrees of freedom is three.

\section{Summary and discussions}
\label{sec:summary}

We have investigated the nature of constraints and the Hamiltonian
structure in the scalar-tensor theory recently proposed by Gleyzes,
Langlois, Piazza and Vernizzi (GLPV)~\cite{Gleyzes:2014dya}. For the
simple case with $A_5=0$, we have proved that the number of independent
degrees of freedom is three at fully nonlinear level, as claimed by
GLPV. 

The Hamiltonian analysis in the present paper is similar to but actually
differs from that done by GLPV for a couple of reasons. First, in the
present paper the momentum constraint, that is the generator of the
spatial diffeomorphism, is given by (\ref{eqn:def-Htot_i}). Compared
with the corresponding expression by GLPV, this includes an additional
term of the form $\pi_N\partial_iN$, where $N$ is the lapse function and
$\pi_N$ is the canonical momentum conjugate to $N$. The presence of the 
additional term of this form is expected from physical viewpoints:
the time-space component of the stress-energy tensor of a scalar field
should contribute to the total momentum constraint but the scalar field
is encoded in the lapse function in the unitary gauge. Indeed, without
the additional term, the Poisson bracket between the momentum constraint
and the other secondary constraint ${\cal C}$ would not vanish weakly,
and thus the momentum constraint would not be first-class. Second, in
the present paper we include not only $A_{2,3,4}$ and $B_4$ but also
$B_5$. In spite of these differences, our analysis still supports the
claim by GLPV: the number of degrees of freedom is three.

It is expected that inclusion of $A_5\ne 0$ and more general
terms~\cite{Gao:2014soa} does not change the constraint algebra and thus
the number of degrees of freedom. However, the analysis becomes
technically involved and thus we consider it as beyond the scope of the
present paper.

The Hamiltonian analysis in the present paper is based on the unitary
gauge, in which the scalar field is encoded in the lapse function (and
the time variable). Extension of the analysis to a general gauge is also
beyond the scope of the present paper but we would like to make some
comments on it here. It should be possible to obtain canonical
transformation that maps the set of phase space variables in the unitary
gauge to that in a general gauge. The constraints among the phase space
variables in the unitary gauge, which we analyzed in the present paper,
are then transformed to those in the general gauge. The algebra of
constraints should be the same in any gauge. Together with the
first-class constraint corresponding to the time diffeomorphism, it
should be possible to show that the dimension of the physical phase
space is six and that the number of degrees of freedom is
three. However, we shall leave this analysis to a future work. 

\section*{\bf Acknowledgments}
 This work was supported in part by WPI Initiative, MEXT, Japan,
 Grant-in-Aid for Scientific Research 24540256 and Grant-in-Aid for JSPS
 Fellows. We would like to thank N.~Tanahashi for useful discussions.

\appendix

\section{Hamiltonian analysis of constrained system}
\label{app:constraints}

In this appendix, we summarize the standard Hamiltonian analysis of a
system with constraints~\footnote{
In this appendix, we consider a finite number of coordinate variables
$q_I$ for simplicity. If we extend this formalism to a (bosonic) field
theory, we consider field variables instead and perform the procedure in
a similar manner, as we have done in the main text. In this case,
however, the coordinate indices $I$ and $J$ in (\ref{action-app}) run
through infinity, and $\int dt L$ is replaced by $\int d^4x {\cal L}$,
where ${\cal L}$ is a Lagrangian density. In principle, it is not
guaranteed that the procedure described in this appendix ends by a
finite number of steps for field theories, for which $N$ is infinite. 
}.
The standard analysis was introduced by P. Dirac in 50s and 60s
\cite{Dirac}, as a way of quantizing mechanical systems such as gauge
theories.

We consider a system which has the following action,
\begin{equation}
    S = \int dt \, L(q_I,\, \dot q_J,\, t)\ ,\qquad I,J= 1,2,...\,,N,
    \label{action-app}
\end{equation}
where $q_I$ are Grassmann even coordinate variables and an overdot
denotes a derivative with respect to time $t$. The Lagrangian $L$ may
depend on time explicitly. The canonical momentum $p^I$ and the
Hamiltonian $\tilde H$ are defined as 
\begin{align}
\label{pi}
   p^I &\equiv \frac{\partial L}{\partial \dot q_I}\ ,\\
   \tilde H &\equiv p^I\dot q_I - L \ .
\end{align}
In the cases where the system is singular, (\ref{pi}) cannot completely be solved for $\dot q_I$; this happens when 
\begin{equation}
\label{det}
\text{det}\left|\frac{\partial^2L}{\partial \dot q_I\partial \dot q_J}\right| = 0\ .
\end{equation} 
Eq.~(\ref{det}) means that we have $M_1 = N-r_0$ constraints independent from $\dot q_I$, where $r_0= \text{rank}(\partial^2 \mathrm{L}/\partial \dot q_I \partial \dot q_J)$. We denote these ``primary'' constraints as
\begin{equation}
\label{phi}
    \phi_A(q_I,\, p^J,\, t)=0 \ ,\qquad A=1,2,...,M_1 \ .
\end{equation}
Including the information of the constraints into the Hamiltonian, we define a new Hamiltonian
\begin{equation}
H = \tilde H + \phi_A\lambda^A \ .
\end{equation} 
where $\lambda^A$ are Lagrange multipliers. The inclusion of $\phi_A \lambda^A$ should not be considered artificial; it merely represents the degree of arbitrariness proportional to $\phi_A$ in defining the Hamiltonian from the Lagrangian.
Thus $\tilde H$ and $H$ cannot be physically distinguished on the surface defined by $\phi_A = 0$. 

For an arbitrary function $F(q_I,\, p^J,\, t)$, we can find its time evolution using the canonical equations by
\begin{equation}
\label{F1}
\frac{d}{dt}F = \frac{\partial}{\partial t}F + \{F,\, \tilde H\}_{\rm P} + \{F,\, \phi_A\}_{\rm P}\lambda^A 
		   \approx \frac{\partial}{\partial t}F + \{F,\, H\}_{\rm P} \ ,
\end{equation}
where the weak equality $\approx$ in (\ref{F1}) holds once the constraints (\ref{phi}) are imposed, and Poisson brackets are defined as
\begin{equation}
\{ F , G \}_{\rm P} \equiv \frac{\partial F}{\partial q_I} \, \frac{\partial G}{\partial p^I} - \frac{\partial F}{\partial p^I} \, \frac{\partial G}{\partial q_I} \ , \qquad {}^\forall \ F(q_I , p^J , t) \; , \; {}^\forall \ G(q_I , p^J , t) \ .
\end{equation}
The term with $\phi_A$ in (\ref{F1}) appears due to the requirement that the variations must be taken on the constraint surface defined by (\ref{phi}).
In order for the constraints (\ref{phi}) to hold throughout the time evolution (on the constraint surface), we require the following consistency conditions; substituting $\phi_A$ into $F$ in (\ref{F1}), we have 
\begin{equation}
\label{con}
   \frac{d}{dt}\phi_A  = \frac{\partial}{\partial t} \phi_A + \{ \phi_A , \tilde{H} \}_{\rm P} + \left\{ \phi_A , \phi_B \right\}_{\rm P} \lambda^B
                               \approx 0 \ .
\end{equation}
While in some cases these conditions merely fix some Lagrange
multipliers $\lambda^A$, in other cases they introduce additional
constraints, called ``secondary'' constraints. To see this, we take the
linear combinations of $\phi_A$ properly (such as done for
$\mathcal{H}_i^{\rm tot}$ in the main text), to reduce the coefficient
matrix $\{\phi_A,\, \phi_B\}_{\rm P}$ in (\ref{con}) to the form
\begin{align}
\label{mat}
&\{\phi_A,\, \phi_B\}_{\rm P} =
\begin{pmatrix}
   \{\phi_\alpha,\, \phi_\beta\}_{\rm P}   &   \{\phi_\alpha,\, \phi_b\}_{\rm P} \\
   \{\phi_a,\, \phi_\beta\}_{\rm P}   &   \{\phi_a,\, \phi_b\}_{\rm P}
\end{pmatrix}
\approx
\begin{pmatrix}
    C_{\alpha \beta}    &  0  \\
     0 &  0
\end{pmatrix} \ , 
\end{align}
where $\det (C) \not \approx 0$, $A, B = 1, \dots, M_1$, 
$\alpha , \beta = 1, \dots , r_1$, $a,b = r_1+1 , \dots , M_1$, and 
$r_1=\mathrm{rank}\left( \{\phi_A,\, \phi_B\}_{\rm P} \right)$. Since
any odd-dimensional antisymmetric matrix has vanishing determinant,
$r_1$ here is always an even number. Since there exists the inverse
matrix of $C_{\alpha \beta}$, $\lambda^\alpha$ can be uniquely
determined by
\begin{equation}
\label{lambda-sec}
\lambda^\alpha = - (C^{-1})^{\alpha \beta}\left( \frac{\partial}{\partial t}\phi_\beta
                             + \{\phi_\beta,\, \tilde H\}_{\rm P}\right) \ .
\end{equation} 
On the other hand, we cannot determine the remaining ($M_1-r_1$) multipliers $\lambda^a$, as
the conditions (\ref{con}) for $\phi_a$ reduce to 
\begin{equation}
\label{2nd}
\frac{d}{dt}\phi_a = \frac{\partial}{\partial t}\phi_a + \{\phi_a,\, \tilde H\}_{\rm P} 
                           \approx \frac{\partial}{\partial t}\phi_a + \{\phi_a,\, H\}_{\rm P} \approx 0\ .
\end{equation}
If $d \phi_a / dt$ can be expressed as linear combinations of $\phi_A$, then no further procedure is necessary; otherwise, however, (\ref{2nd}) will introduce the ``secondary'' constraints.
If we obtain $M_2$ secondary constraints , we combine them with the primary 
constraints and extend the indices in (\ref{con}) to $A,B = 1 , \dots , M_1+M_2$.
Then we repeat the steps (\ref{mat} - \ref{2nd}) for the new set of constraints. Eq.~(\ref{2nd}) may again introduce further $M_3$ secondary constraints. Repeating this procedure until (\ref{2nd}) produces no further constraint equations, we finally obtain $M=M_1+M_2+\cdots(\leq2N)$ constraints and the coefficient matrix $\{\phi_A, \phi_B\}_{\rm P}$ $(A,\, B = 1 , \dots , M )$ whose rank is $r = r_1+r_2+\cdots(\leq M)$.

Adding the constraint terms, the ``total'' Hamiltonian is of the form
\begin{equation}
\label{}
H_\mathrm{tot} = \tilde H + \phi_\alpha \lambda^\alpha + \phi_a\lambda^a  \ ,
\end{equation}
where $\lambda^\alpha$ are determined by (\ref{lambda-sec}) (now $\alpha$ runs from $1$ through $r$).
The remaining $(M-r)$ multipliers $\lambda^a$ are yet to be determined. 
In order to fully determine them, we first define a useful terminology 
to distinguish $\phi^a$ from $\phi^\alpha$. We call any dynamical
variable $\mathcal{R}(q_I,\, p^J,\, t)$ \textit{first-class} if
$\mathcal{R}$ satisfies 
\begin{equation}
\label{}
\{\mathcal{R},\, \phi_A\}_{\rm P} \approx 0 \ ,\qquad A=1,2,...,M \ ,
\end{equation}
and otherwise we call it \textit{second-class}. This definition is a slight extension of 
original Dirac's one to the dynamical variables which can depend on time explicitly. 
According to this definition, the constraints $\phi^a$ are first-class, and $\phi^\alpha$ are second-class. 
As we have seen, the system contains the same number of undetermined coefficients $\lambda^a$ as that of the first-class constraints $\phi_a$. This in fact implies that $\phi_a$ are generators of gauge transformation of the system, under which all physical quantities must be invariant.
The number of first-class constraints, $(M-r)$, is equal to the number of gauge symmetry, which in our case is spatial diffeomorphism.
As gauge fixing, we can by hand impose additional $(M-r)$ constraints,
\begin{equation}
\label{gauge-fix}
\chi_a(q_I,\, p^J,\, t) \approx 0 \ .
\end{equation} 
We note that these gauge fixing conditions do not affect the second-class
constraints since gauge symmetry does not change physics (or
mathematically $\{ \phi_a , \phi_\alpha \}_{\rm P} \approx 0$). 
Then, we require the consistency conditions for $\chi_a$ as in (\ref{con}), 
\begin{equation}
\label{consis-gauge}
\frac{d}{dt}\chi_a \approx \frac{\partial}{\partial t}\chi_a + 
                                        \{\chi_a,\, \tilde H + \phi_\alpha \lambda^\alpha\}_{\rm P} +
                                        \{\chi_a,\, \phi_b\}_{\rm P}\lambda^b \approx 0 \ .
\end{equation}
The choice of the gauge fixing conditions (\ref{gauge-fix}) should not be completely arbitrary, but rather they are to determine, through (\ref{consis-gauge}), the remaining Lagrange multipliers $\lambda^a$, equivalently fixing the gauge completely.
Therefore we require,
\begin{equation}
\mathrm{det} \left\vert \{\chi_a,\, \phi_b\}_{\rm P} \right\vert \not \approx 0 \ ,
\end{equation} 
which leads to the relation,
\begin{equation}
\mathrm{det}\left|\begin{array}{cc}
     \{\chi_a,\, \chi_c\}_{\rm P} & \{\chi_a,\, \phi_d\}_{\rm P}   \\
     \{\phi_b,\, \chi_c\}_{\rm P} & \{\phi_b,\, \phi_d\}_{\rm P} 
\end{array}\right|
 \approx \mathrm{det}^2\left|\{\chi_a,\, \phi_b\}_{\rm P}\right| \not \approx 0 \ .
\end{equation}
given $\{\phi_b,\, \phi_d\}_{\rm P}\approx 0$. Hence $\phi_a$, together
with $\chi_a$, can now be treated as second-class constraints, and we
can determine the remaining $\lambda^a$ and the new Lagrange multipliers
associated with the gauge-fixing constraints (\ref{gauge-fix}). We have
therefore shown that once we fix the gauge completely and determine all
the multipliers, we can solve for the motion of the system in that
gauge, at least classically.

\section{Outline of calculating Poisson brackets}
\label{app:poisson}

In the main text, we spared all the detailed calculations of the Poisson brackets and focused on the Hamiltonian structure of the theory. In this appendix, we outline some of the omitted part of the calculations. Among the Poisson brackets we have computed, the only non-trivial ones are $\left\{ \bar{\cal H} , \bar{\cal H}\right\}_{\rm P}$ and $\left\{ \bar{\cal H} , \bar{\cal C} \right\}_{\rm P}$ in (\ref{HbarHbar}) and (\ref{HbarCbar}), respectively (in principle, one may consider $\left\{ \bar{\cal C} , \bar{\cal C} \right\}_{\rm P}$, but there is no need to compute it in order to study the structure of the theory). 

First it is useful to know the relation
\begin{equation}
\sqrt{h} \, D_i V^i = \partial_i \left( \sqrt{h} \, V^i \right) \; ,
\label{cov2par}
\end{equation}
where $V^i$ is an arbitrary vector. Thus if the expression on the
left-hand side of (\ref{cov2par}) appears in the $3$-integral $\int
d^3x$, then it becomes total derivative. The relation (\ref{cov2par}) is
used throughout the paper, whenever applicable. 

The variations of $\bar{\cal H}$ and $\bar{\cal  C}$ with respect to $h_{ij}$ and $\pi^{ij}$ are
\begin{eqnarray}
\frac{\delta \bar{\cal H} [f]}{\delta h_{ij} (z)} & \approx & \sqrt{h} \left[ \frac{\pi^{il}}{\sqrt{h}} D_l f^j + \frac{\pi^{jl}}{\sqrt{h}} \, D_l f^i - D_l \left( f^l \frac{\pi^{ij}}{\sqrt{h}} \right) \right]  \; ,
\label{Hbar-hij}\\
\frac{\delta \bar{\cal H} [f]}{\delta \pi^{ij} (z)} & \approx & D_i f_j + D_j f_i  \; ,
\label{Hbar-pij}\\
\frac{\delta \bar{\cal C} [\varphi]}{\delta h_{ij} (z)} & \approx & \frac{1}{2} \varphi \, {\cal C} h^{ij} + \sqrt{h} \Bigg\{
\varphi \, \frac{\pi^2 h^{ij} - 2 \pi_{mn} \pi^{mn} h^{ij} - 2 \pi \pi^{ij} + 4 \pi^i_{\;\; l} \pi^{lj}}{2 h} \, \frac{\partial}{\partial N}\left( \frac{N}{A_4} \right)
+ \varphi \, \frac{2 \pi^{ij} - \pi h^{ij}}{4 \sqrt{h}} \, \frac{\partial}{\partial N} \left( \frac{N A_3}{A_4} \right)
\nonumber\\
&&
- \varphi \, R^{ij} \frac{\partial}{\partial N} \left( N B_4 + \frac{N A_3 B_5}{8 A_4} \right)
+ \varphi \, \frac{2 \pi^{ij} R - 2 \pi R^{ij} + 4 \pi^{mn} R_{mn} h^{ij} - \pi R h^{ij}}{8 \sqrt{h}} \, \frac{\partial}{\partial N} \left( \frac{N B_5}{A_4} \right)
\nonumber\\
&&
- \frac{\varphi}{2} \left( R^i_{\;\; l} R^{jl} - \frac{3}{8} R \, R^{ij} \right) \frac{\partial}{\partial N} \left( \frac{N B_5^2}{A_4} \right)
+ \left( D^i D^j - h^{ij} D^2 \right) \left[ \varphi \, \frac{\partial}{\partial N} \left( N B_4 + \frac{N A_3 B_5}{8 A_4} \right) \right]
\nonumber\\
&&
+ \frac{1}{4} \left[ \delta^l_k \left( D^i D^j - h^{ij} D^2 \right) + 2 \left( \delta^i_k h^{jl} D^2 + h^{ij} D_k D^l - \delta^i_k D^l D^j - \delta^j_k D^l D^i  \right) \right]
\nonumber\\
&& \qquad
\times \left[ \varphi \, \frac{\pi^k_{\;\; l}}{\sqrt{h}} \, \frac{\partial}{\partial N} \left( \frac{N B_5}{A_4} \right) - \frac{\varphi}{2} G^k_{\;\; l} \, \frac{\partial}{\partial N} \left( \frac{N B_5^2}{A_4} \right) \right]
\Bigg\} \; ,
\label{C-hij}\\
\frac{\delta \bar{\cal C} [\varphi]}{\delta \pi^{ij} (z)} & \approx & \varphi \Bigg\{
\frac{2 \pi_{ij} - \pi h_{ij}}{\sqrt{h}} \, \frac{\partial}{\partial N} \left( \frac{N}{A_4} \right)
+ \frac{1}{2} h_{ij} \, \frac{\partial}{\partial N} \left( \frac{N A_3}{A_4} \right) 
- \left( G_{ij} + \frac{1}{4} R \, h_{ij} \right) \frac{\partial}{\partial N} \left( \frac{N B_5}{A_4} \right)
\Bigg\} \; ,
\label{C-pij}
\end{eqnarray}
where $D^i \equiv h^{ij} D_j$, $D^2 \equiv h^{ij} D_i D_j$, $f_i \equiv h_{ij} f^j$, $G_{ij} \equiv R_{ij} - \frac{1}{2} R h_{ij}$, and the weak equality $\approx$ implies that the equality holds if $f^i$ and $\varphi$ do not depend on $h_{ij}$ or $\pi^{ij}$. 
In order to derive (\ref{C-hij}), we have used the variations of $R_{ij}$ and $R$, which are given by %
\footnote{
From (\ref{R-var}), it is immediate to see
\begin{equation}
\delta ( \sqrt{h} R )  = \sqrt{h} \left( - G^{ij} + D^j D^i - h^{ij} D^2 \right) \delta h_{ij} \; ,
\nonumber
\end{equation}
as expected. Note that the last two terms in the parentheses would be total derivatives if $\sqrt{h} R$ appears in the action by itself (up to constant coefficients).
}
\begin{eqnarray}
\delta R_{ij} & = & \frac{1}{2} h^{kl} D_l \left( D_i \, \delta h_{jk} + D_j \, \delta h_{ik} - D_k \, \delta h_{ij} \right) - D_j D_i \, \delta \ln \sqrt{h} \; , \\
\delta R & = & \left( - R^{ij} + D^j D^i - h^{ij} D^2 \right) \delta h_{ij} \; . \label{R-var}
\end{eqnarray}
In principle $\bar{\cal C}$ has non-vanishing variation with respect to $N$ as well, but it is not needed for the current purpose, since $\frac{\delta \bar{\cal H}}{\delta \pi_N} = 0$.

Using (\ref{Hbar-hij}) and (\ref{Hbar-pij}), it is straightforward to show
\begin{equation}
\left\{ \bar{\cal H} [f] , \bar{\cal H} [g] \right\}_{\rm P} \approx \int d^3x \Bigg\{
2 \sqrt{h} D_i \left[ \epsilon^{ijk} \epsilon_{mnk} \left( f^n D_j g^l - g^n D_j f^l \right) \frac{\pi^m_{\;\;\; l}}{\sqrt{h}} \right]
- 2 D_j \sqrt{h} \left( \frac{\pi^{j}_{\;\; i}}{\sqrt{h}} \right) \left( f^j D_j g^i - g^j D_j f^i \right)
\Bigg\} \; .
\label{HH-app}
\end{equation}
The term with the square brackets in (\ref{HH-app}) has the structure of (\ref{cov2par}) and thus is total derivative. Hence we find $\left\{ \bar{\cal H} [f] , \bar{\cal H} [g] \right\}_{\rm P} \approx \bar{\cal H} \left[ [f,g] \right]$, as in (\ref{HbarHbar}).

The calculation of $\left\{ \bar{\cal H} , \bar{\cal C} \right\}_{\rm P}$ is more involved, yet straightforward. 
For the ease of the calculation, we remind of some properties of curvature tensors. The Bianchi identity with some indices contracted is often found useful; in particular,
\begin{eqnarray}
D_l R_{ij} - D_i R_{lj} & = & D_k R^k_{\;\; jli} \; , \\
D_i G^i_{\; j} & = & 0 \; .
\end{eqnarray}
The commutator of covariant derivatives introduces the curvature tensors; when acting on an arbitrary tensor, it is
\begin{eqnarray}
\left[ D_i \, , D_j \right] X^{m_1 \dots m_a}_{\qquad\;\; n_1 \dots n_b} 
& = & 
R^{m_1}_{\;\;\;\; kij} X^{k m_2 \dots m_a}_{\qquad\;\;\; n_1 \dots n_b} + \dots
+ R^{m_a}_{\;\;\; kij} X^{m_1 \dots m_{a-1} k}_{\qquad\quad\;\;\; n_1 \dots n_b}
\nonumber\\
&& \!\!\!\!\!
- R^k_{\;\; n_1 ij} X^{m_1 \dots m_a}_{\qquad\;\; k n_2 \dots n_b} - \dots
- R^k_{\;\; n_b ij} X^{m_1 \dots m_a}_{\qquad\;\; n_1 \dots n_{b-1} k} \; ,
\label{com-covder}
\end{eqnarray}
since the connection in the present case is torsion free. Using these
relations, one finds, up to total derivatives, 
\begin{eqnarray}
\left\{ \bar{\cal H} [f] , \bar{\cal C} [\varphi] \right\}_{\rm P} & \approx & \int d^3x \sqrt{h} \Bigg\{
\left( \frac{\pi^i_{\;\; j} \pi^j_{\;\; i}}{h} - \frac{\pi^2}{2 h} \right) f^i D_i \left[ \varphi \, \frac{\partial}{\partial N} \left( \frac{N}{A_4} \right) \right]
+ \frac{\pi}{2 \sqrt{h}} f^i D_i \left[ \varphi \, \frac{\partial}{\partial N} \left( \frac{N A_3}{A_4} \right) \right]
\nonumber\\
&&
- \frac{3}{8} f^i D_i \left[ \varphi \, \frac{\partial}{\partial N} \left( \frac{N A_3^2}{A_4} \right) \right]
+ f^i D_i \left[ \varphi \, \frac{\partial}{\partial N} \left( N A_2 \right) \right]
+ R \, f^i D_i \left[ \varphi \, \frac{\partial}{\partial N} \left( N B_4 \right) \right]
\nonumber\\
&&
\nonumber\\
&&
- \frac{\pi^{ij} R_{ij} - \frac{1}{4} \pi R}{\sqrt{h}} f^i D_i \left[ \varphi \, \frac{\partial}{\partial N} \left( \frac{N B_5}{A_4} \right) \right]
+ R \, f^i D_i \left[ \varphi \, \frac{\partial}{\partial N} \left( \frac{N A_3 B_5}{8 A_4} \right)\right]
\nonumber\\
&&
+ \left( R^m_{\;\; n} R^n_{\;\; m} - \frac{3}{8} R^2 \right) f^i D_i \left[ \varphi \, \frac{\partial}{\partial N} \left( \frac{N B_5^2}{4 A_4} \right) \right]
\Bigg\} \; .
\end{eqnarray}
We can rewrite this expression to be
\begin{equation}
\left\{ \bar{\cal H} [f] , \bar{\cal C} [\varphi] \right\}_{\rm P} \approx \int d^3x \left( {\cal C} f^i D_i \varphi + \varphi  \, f^i D_i N \, \frac{\partial}{\partial N} {\cal C}\right) \; .
\label{Hbar-Cbar}
\end{equation}
The first term vanishes on the surface defined by ${\cal C} = 0$, but the second term does not. In order for ${\cal H}_i$ to be first-class constraints, therefore, we need to introduce another term in ${\cal H}_i$ to have
\begin{equation}
{\cal H}_i^{\rm tot} \equiv {\cal H}_i + \pi_N \partial_i N \; ,
\end{equation}
as in the main text. This new term cancels out the non-vanishing term in (\ref{Hbar-Cbar}), giving
\begin{equation}
\left\{ \bar{\cal H}^{\rm tot} [f] , \bar{\cal C} [\varphi] \right\}_{\rm P} \approx \bar{\cal C} [ f \partial \varphi] \; ,
\end{equation}
which concludes the proof of the calculations of the Poisson brackets in analyzing the Hamiltonian structure of the theory.

\end{document}